# DETECTION OF METALLO-β-LACTAMASES-ENCODING GENES AMONG CLINICAL ISOLATES OF ESCHERICHIA COLI IN A TERTIARY CARE HOSPITAL, MALAYSIA

FAZLUL MKK[1], DEEPTHI S[2], FARZANA Y[3], NAJNIN A[4], RASHID MA[5], MUNIRA B[6], SRIKUMAR C.[2], *NAZMUL

[1]Faculty of Industrial Sciences & Technology, University Malaysia Pahang, Gambang, 26300 Pahang, Malaysia
[2]Graduate School of Medicine, Perdana University, Jalan MAEPS Perdana, Serdang, 43400 Selangor, Malaysia
[3]Faculty of Science, Lincoln University, 12-18, Jalan SS6/12, Off Jalan Perbandaran, 47301 Petaling Jaya, Selangor Malaysia.
[4]Jeffrey Cheah School of Medicine and Health Sciences, Monash University, No.8, Jalan Masjid Abu Bakar, 80100 Johor Bahru, Malaysia
[5]Faculty of Medicine, University Teknologi MARA, Jalan Hospital, Sg Buloh, Selangor 47000, Malaysia
[6]Faculty of Medicine, SEGi University, Kota Damansara, Selangor, Malaysia
Email id : poorpiku@yahoo.com
Received: 05.11.18, Revised: 21.01.19, Accepted: 26.02.19

**ABSTRACT**

The multidrug resistant *Escherichia coli* strains causes multiple clinical infections and has become a rising problem globally. The metallo-β-lactamases encoding genes are very severe in gram-negative bacteria especially *E. coli*. This study was aimed to evaluate the prevalence of MBLs among the clinical isolates of *E. coli*. A total of 65 *E. coli* isolates were collected from various clinical samples of Malaysian patients with bacterial infections. The conventional microbiological test was performed for isolation and identification of *E. coli* producing MBLs strains in this vicinity. Multidrug Resistance (MDR) of *E. coli* isolates were assessed using disk diffusion test. Phenotypic methods, as well as genotypic- PCR methods, were performed to detect the presence of metallo-β-lactamase resistance genes (blaIMP, blaVIM) in imipenem resistant strains. Out of 65 *E. coli* isolates, 42 isolates (57.3%) were MDR. The isolates from urine (19) produced significantly more MDR (10) isolates than other sources. Additionally, 19 (29.2%) imipenem-resistant *E. coli* isolates contained 10 MBLs gene, 7(36.8%) isolates contained blaIMP and 3(15.7%) isolates contained blaVIM genes. This study revealed the significant occurrence of MBL producing *E. coli* isolates in clinical specimens and its association with health risk factors indicating an alarming situation in Malaysia. It demands an appropriate concern to avoid failure of treatments and infection control management.

## INTRODUCTION

Treatment of bacterial infections has become complicated due to the emergence of multi-drug resistant strains of *Escherichia coli*. *E. coli* showed the highest antibiotic resistance trends among the different types of bacteria in Malaysia between 2013 and 2017 (Fazlul *et al.*, 2018). Metallo-beta-lactamases (MBLs) are beta-lactamase enzymes produced by pathogenic bacteria and gradually found in Gram-negative organisms, mostly in *E. coli* species. Clinical infections with MBL-producing isolates are associated with higher morbidity and mortality (Deshmukh *et al.*, 2011). However, carbapenem resistance due to Metallo-beta-lactamases (MBLs) production has been gradually reported among clinical isolates of *E. coli* (Bora *et al.*, 2014; Nepal *et al.*, 2017). Among *Enterobacteriaceae* species, the blaIMP and blaVIM genes have been identified throughout the world. The prevalence rate of carbapenem resistance *E. coli* is higher by acquiring MBL including imipenem (IPM) and Verona integron-encoded Metallo-β-lactamase (VIM) around the world. Metallo-β-lactamases (MBLs) which hydrolyze the carbapenems (imipenem, meropenem, and ertapenem) and render them ineffective for treatment (Meletis, 2016; Morrill *et al.*, 2015). In addition, MBLs are not susceptible to therapeutic β-lactamase inhibitors like sulbactam, tazobactam, or clavulanic acid (Drawz *et al.*, 2014; Everaert and Coenye, 2016; Watkins *et al.*, 2013). The emergence and uncontrolled spread of carbapenems in Gram-negative bacteria are under threat (Chika *et al.*, 2014) but normally carbapenems are used for the treatment of infections caused by β-lactam resistant bacteria including extended-spectrum enzymes producers (Bashir *et al.*, 2011). The unusual reduced susceptibility of bacterial pathogens to MBLs as reported as increasing concern matter (Chakraborty et al., 2010). Carbapenem-resistant *Enterobacteriaceae* can cause several severe infections, especially in healthcare settings over the last decade. The rising carbapenems resistant organisms has become a risk factor to existing antibiotics used for handling nosocomial infections (e.g. bacteremia, septicemia, and pneumonia in children). The recurrence rate of five different types of MBLs (IMP, VIM, SPM, GIM and SIM) are increasing rapidly while IMP and VIM types existence in-universe (Gupta, 2008). In vitro chelating agents (EDTA and dipicolinic acid) inhibit MBLs enzyme, can be detected phenotypically in pathogenic bacteria (Chika et al., 2014). MBLs gene expression can be either plasmid-mediated or chromosomal (El Salabi et al., 2013; Smet et al., 2010). The standard methods are modified Hodge test (MHT), double-disc synergy test (DDST), combined disc diffusion test using imipenem and EDTA, and MBL E-test (Sachdeva et al., 2017). Among the phenotypic methods, DDST methods are less sensitive compared to CDT for detecting MBL genes (Jorth et al., 2017; Mehta and Prabhu, 2016; Nermine et al., 2015) but earlier MHT were much sensitive then DDST (Shivaprasad et al., 2014). MBLs production can be assessed by molecular methods and phenotypic methods. Among the molecular techniques, polymerase chain reaction (PCR), DNA probes, cloning and sequencing are highly accurate to identify MBL-positive genes (Sachdeva et al., 2017). Metallo beta-lactamase producing gram-negative bacteria are frequently shown resistance to variance classes of drugs associated with multi-drug resistant bacteria (Meletis, 2016). Hence, the consistent detection of the MBL-producing genes is vital for appropriate treatment for nosocomially infected patients and quality management of multi-drug resistant bacteria (Doosti et al., 2013). The improvement and innovation of antimicrobials agents with novel innovation are in need to manage infectious diseases effectively (Fazlul MKK, 2018).The objective of this study was to evaluate the prevalence rate of MBLs producing *E. coli* and its increasing trends in Malaysia. The clinical observation of this study will be helpful for rapid detection of resistance mechanism due to seveal metallo- -lactamases-encoding genes to prevent the spreading of these microorganisms.

**Materials And Methods**

Sixty-five strains of bacterial samples were collected and identified as *Escherichia coli* using 16SrRNA with the universal primers 27f (5'-AGAGTTTGATCCTGGCTCAG-3') and 1492r (5'-GGTTACCTTGTTACGACTT-3') (Tissera and Lee, 2013). Only the patients having bacteriological evidence of clinical features were included in this study. The sources of studied isolates are as follows: urine, 19 (29%); wounds, 9(14%); respiratory tract, 12(18%); stool, 17(26%); sputum, 3(5%) and blood, 5(8%).

**Antimicrobial susceptibility Test**

Antibiotic susceptibility was performed for all 65 clinical isolates of Escherichia coli by Kirby Bauer disc diffusion method in accordance with CLSI-2017 guidelines incorporating standard strain of Escherichia coli. Different types of antibiotics (tazobactam 10/ piperacillin 75- TZP, ceftazidime-CAZ, ciprofloxacin-CIP, meropenem-MEM, imipenem-IPM, cefepime-FEP, cefotaxime-CTX, aztreonam-ATM, amoxicillin-clavulanic acid-AMC, cefoperazone -CPZ) were used for susceptibility patterns of E. coli. In the screening of MBL producing E. coli strains, EDTA powder was used. Carbapenem-resistant were considered when the zone of inhibition was around imipenem ≤ 13 mm, intermediate 14-15 mm and sensitive ≥ 16 mm (CLSI, 2017). Imipenem-resistant isolates were further screened for of metallo- -lactamases-encoding genes.

**Modified Hodge test**

The Modified Hodge Test (MHT) was performed to detect carbapenemase among the Escherichia coli strains in this study. A lawn culture of 1:10 dilution of 0.5 McFarland's standard Escherichia coli ATCC 25922 was cultured on a Muller-Hinton (MH) agar and allowed to dry 3-5 minutes. A 10 µg imipenem disc was placed in the center of the plate. Imipenem-resistant E. coli (test isolates) in a straight line, streaked from the edge of the disk to the edge of the plate. Four E. coli strains were tested on the same plate with one drug. The carbapenem disc of imipenem (10 $\mu$g) or meropenem (10 $\mu$g) was placed at the middle of the plate. E. coli growth was spotted in imipenem disk (within the inhibition zone) by indicating a distorted zone.

**The double-disc synergy test**

Double disk synergy test was assessed by the CLSI guidelines for the disk diffusion method (CLSI, 2017). An opacity of 0.5 McFarland of test organism was streaked on the MHA using a sterile cotton swab to get a lawn culture. Imipenem (10 ug) disc was placed in 20 mm center to center from a blank disc contained 10 $\mu$l of 0.5 M EDTA (750 ug) or EDTA disc (750 ug). The cultured plate was kept in incubator for 16 to 18 hours at 37°C (Fazlul MKK, 2018; Nazmul et al., 2017; Suryadevara et al., 2017). The occurrence of a synergistic inhibitory zone was viewed indicating MBL positive isolates.

**Combined disc diffusion method**

A lawn culture of test isolates (0.5 McFarland opacity standard) was streaked on MH agar and left it to dry for 3-5min. Two 10 µg imipenem discs were kept on inoculated plates at distance of 25mm. Ten µl of 0.5 M EDTA solution was added to one of the imipenem discs. The zone of inhibition of imipenem + EDTA discs compared to imipenem alone is >7 mm after overnight incubation. The test outcome was considered as positive (Kali et al., 2013; Suryadevara et al., 2017) and the procedure was repeated twice to ensure the results.

### E-test

An E-test MBL strip contains imipenem (4–256 µg/ml) and imipenem (1–64 µg/ml) in combination with a fixed concentration of EDTA. A 0.5 McFarland of test isolates was cultured on on MH agar. E-test strip was inoculated for 24 hours on the surface of the agar, and the plates were observed for imipenem and imipenem–EDTA minimum inhibitory concentration (MIC) values more than eight was considered as positive

### DNA preparation and PCR amplification

DNA template preparation was performed to detect the MBLs resistant genes of E. coli isolates and suspended in 300 µL of water (sterile distilled) and boiled for 10 min. Then centrifuge at 12000 × g for 10 min. Formation of supernatant was used as a template DNA for PCR amplification of blaIMP and blaVIM genes. PCR amplification was solution containing 200 $\mu$M concentrations of dNTP, of each primer (10 pM), 1.5 mM MgCl2, Taq polymerase (0.5 U) and 50 ng DNA templates of total 25 $\mu$L final volume. The amplicon sizes and sequences of primers are shown in Table 1. The blaIMP and blaVIM genes were amplified under a specific condition (Table 2). Escherichia coli ATCC 25922 was used as negative controls. The PCR products were visualised by agarose gel electrophoresis stained with ethidium bromide (5 µg/100 mL).

Table 1: Primers for PCR amplification

| Gene target (s) | Primer sequence | Expected amplicon size (bp) | References |
|---|---|---|---|
| blaIMP | 5'-CTACCGCAGCAGAGTCTTTG-3' <br> 5'-AACCAGTTTTGCCTTACCAT -3' | 587 | (Shibata et al., 2003) |
| blaVIM | 5'-CCGATGGTGTTTGGTCGCAT -3' <br> 5'-GAATGCGCAGCACCAGGA -3' | 261 | (Athanassios Tsakris et al., 2000) |

Table 2: PCR programs for amplification

| Target gene | First denaturation | Extension | Annealing | Denaturation | Final extension | References |
|---|---|---|---|---|---|---|
| blaVIM | 95°C 5 min | 72°C 7 min | 72°C 1 min | 58°C 1 min | 95°C 1 min | (Yousefi et al., 2010) |
| blaIMP | 95°C 5 min | 72°C 10 min | 72°C 1 min | 54°C 40 s | 95°C 1 min | |

Primers were designed to detect the presence of the metallo- -lactamase genes, blaVIM (261bp) (A. Tsakris et al., 2000) and blaIMP (587bp) (Shibata et al., 2003) (Table 1). Thermocycling conditions for PCR amplification are shown in table 2.

### Results and Discussion

During this study, antibiotic resistance rate of E. coli isolates (65) was determined. All the E. coli isolates were collected from both male and female patients according to age. The highest 40% of the patients were at the age of 31-45 years old whereas the least (1%) above 60 years (Table 3).

Table 3: Sex-dependent E. coli isolates

| The range of Age Groups in Years | Gender | | Total no (%) |
|---|---|---|---|
| | Male | Female | |
| 1-15 years | 3 | 2 | 8 |
| 16-30 years | 7 | 13 | 31 |
| 31-45 | 17 | 9 | 40 |
| 46-60 | 8 | 5 | 20 |
| Above 60 | 1 | 0 | 1 |

All the clinical strains of E. coli (65) were identified and confirmed by conventional bacteriological tests (16SrRNA) using the universal primers with an expected amplicon of 1500bp (Tissera and Lee, 2013). All E. coli isolates (65) were collected from different clinical samples, urine (29%) followed by stools (26%), respiratory (18%), wounds (14%), blood (8%) and sputum (5%) respectively.Characterisation of Escherichia coli (65) was assessed based on the standard guidelines where E. coli isolates produce pinkish colonies on MacConkey agar, ferments lactose and metallic sheen colonies on Eosin methylene blue agar (Table 4).

**Table 4: Isolation and characterisation of *Escherichia coli***

| Organism | Indole test | Methyl red test | Gram staining | Colonies features | No of Isolates |
|---|---|---|---|---|---|
| *Escherichia coli* | +Ve | +Ve | -Ve | Metallic sheen colonies on EMB and Pinkish colonies on MAC | 65 |

%: Percentage, +Ve: Positive, -Ve: Negative, EBM: Eosin methylene blue agar, MAC: MacConkey agar

Ten different types of commonly prescribed antibiotics have been used in this study to evaluate and to compare the antibiotic susceptibility rate between the isolates from male and female patients but exposed no significant variance ($P > .05$). Sixty-five isolates of *Escherichia coli* clinical isolates were highly resistant to cefalexin, cefepime, aztreonam, and cefotaxime with a resistance rate of 89.2%, 86.1%, 83.6%, and 78.4%, respectively. The least resistant 29.2% to imipenem was observed (Figure 2).

**Table 5: Multidrug-resistant (MDR) among the *E. coli* (65) isolates**

| No and Types of Antibiotics | Types of antibiotics | No (%) of MDR *E. coli* isolates |
|---|---|---|
| 3 classes of antibiotics | CTX, TZP, IMP | 12 (19%) |
| 4 classes of antibiotics | ATM, MEM, CIP | 15 (23%) |
| 5 classes of antibiotics | MEM, AMC, FEP, AMP, CEX | 10 (15%) |
| 6 classes of antibiotics | AMC, CTX, ATM, IMP, CIP, CEX | 3 (5%) |
| 10 classes of antibiotics | TZP, CAZ, CIP, MEM, IMP, FEP, CTX, ATM, AMC, CEX | 2 (3%) |
| | | 42 (65%) |

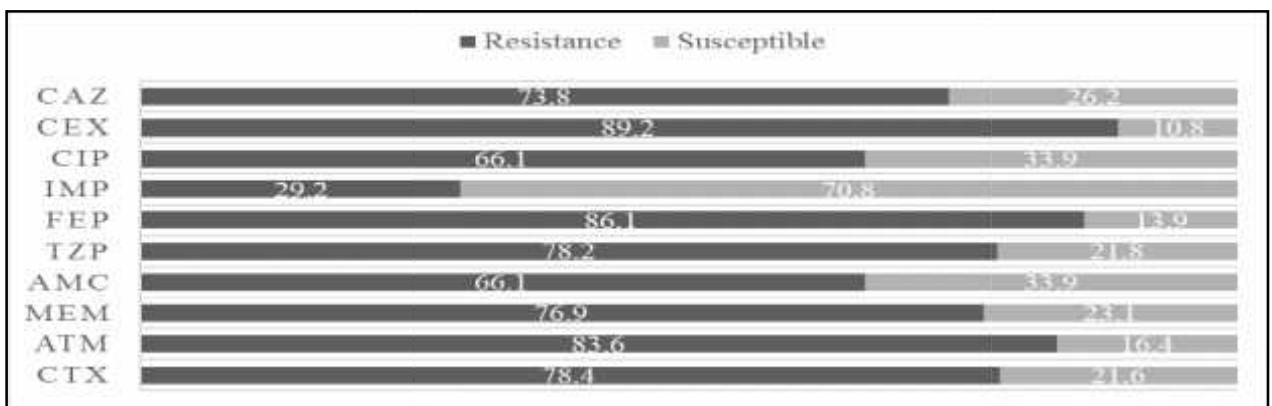

**Figure 1: Antibiotic susceptibility**

Tazobactam 10/ Piperacillin 75- TZP, Ceftazidime-CAZ, Ciprofloxacin-CIP, Meropenem-MEM, Imipenem-IPM, Cefepime-FEP, Cefotaxime -CTX, Aztreonam – ATM, Amoxicillin-clavulanic acid – AMC, Cefalexin -CEX. The prevalence rate of antibiotic resistance among 65 *E. coli* isolates 42 (65%) were found to be resistant to three to six antibiotics and regarded as multi-drug resistant isolates (MDR) (Table 5). Also, 3% (2/65) isolates were resistant to all used antibiotics in this study. Among 42 MDR clinical samples, stool produced the highest resistant organisms 30.9%, followed by respiratory 21.4% and the least was produced by sputum 4.7% (Figure 2).

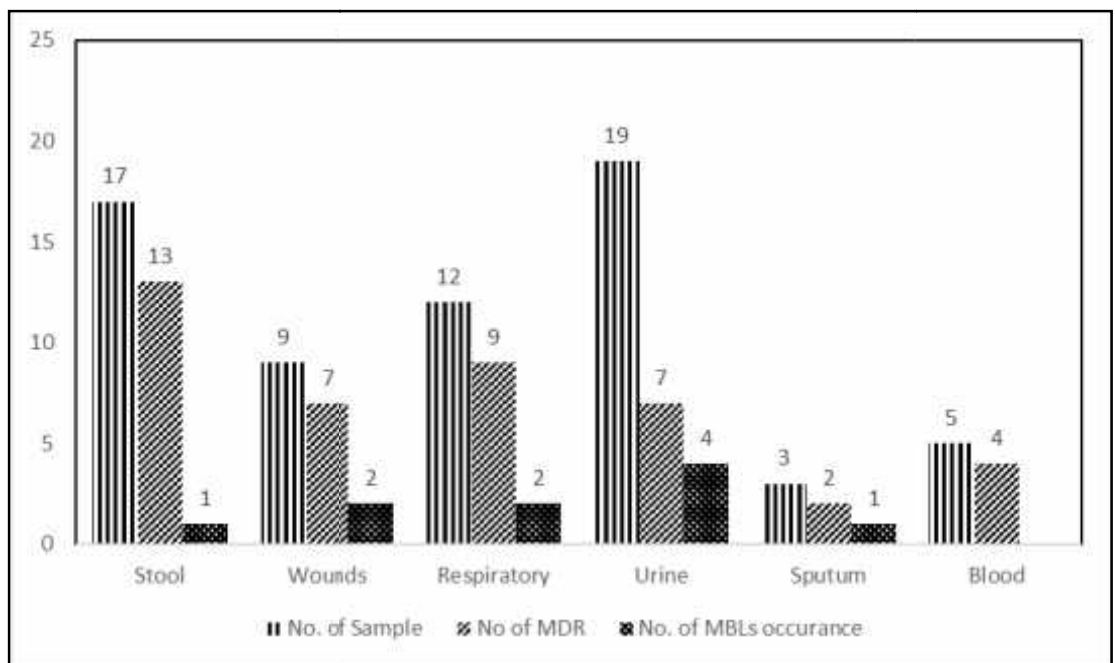

**Figure 2: Frequency of MDR and MBL producers among the clinical specimens**

In this present study, disk diffusion test confirmed 19 (29.2%) isolates resistant to IPM. Among the 19 IPM resistant strains, 12 (63.15%) strains were MIC≥4 µg/ml for IPM by the E-test (bioMérieux, Marcy l'Etoile, France). Phenotypic assays such as CDT, DDST, and MHT were found to be sensitive to determine MBLs producing gene in this study. Among the 19 isolates of imipenem resistant *E. coli*, each of the phenotypic assays differs while the most specific by DDST 52.6%, followed by CDT (47.4%) and MHT was the least specific (36.8%) (Table 6). A zone of >4 mm diameter observed around the imipenem-EDTA disc compared to imipenem disc alone (Figure 3). Studies have shown that phenotypic methods have specificities as similar to PCR and sensitive enough to detect MBL producing isolates (Yalda Khosravi *et al.*, 2012).

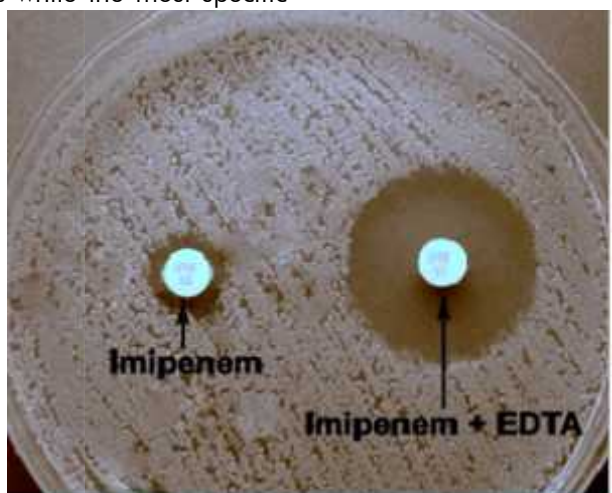

**Figure 3: Positive combined disc test for detection of MBL producer by use of EDTA.**

**Table 6: Frequency of MBL-producing *E. coli* detected by the inhibition-based assay**

| Imipenem-resistant *Escherichia coli* isolates | MHT (Modified Hodge test) | | Double-disc synergy test Ceftazidime + (Ceftazidime + EDTA) E - strip method | | Combined disc test Imepenem + (Imepenem + EDTA) Combined disc diffusion | |
|---|---|---|---|---|---|---|
| | Positive | Negative | Positive | Negative | Positive | Negative |
| 19 | 7(36.8%) | 12(63.2%) | 10 (52.6%) | 9 (47.4%) | 9(47.4%) | 10 (52.6%) |

Imipenem resistance 29.2% *E. coli* strains were screened to detect MBLs gene. MBLs genes (blaIMP, blaVIM) were detected in 10 (52.6%) isolates by PCR genotypic confirmative test. Among these MBLs (10) gene, urine produced the highest 4 (40%) followed by respiratory and wounds 2 (20%) (Figure 1). These 10 strains were then analyzed by PCR and blaIMP gene was present in 7 (12.5%) isolates while 3 (5.4%) isolates carried the blaVIM gene (Figure 4).

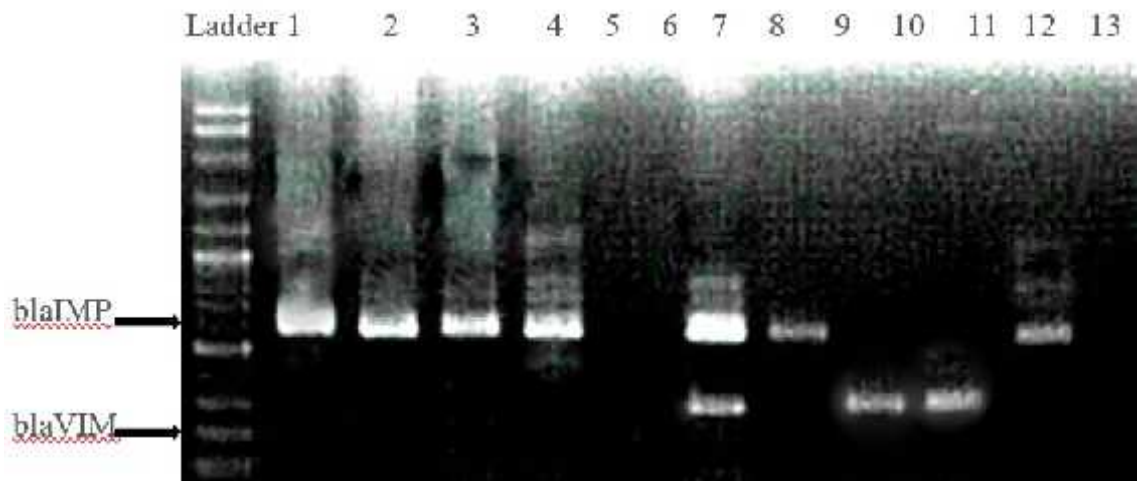

Figure 4: the frequency of MBL genes detected in the 10 *Escherichia coli*

**Discussion**

*Enterobacteriaceae* produces Metallo beta-lactamase (MBL), is one of the mechanisms associated with severe bacterial infections. Antibacterial resistance micro-organisms limit and complicate treatment which is difficult to overcome due to multidrug resistant properties. The outcome of this study showed a high prevalence of *Escherichia coli* from several sources of clinical samples associated with multiple resistance genes. In a previous investigation, sex-dependent antibiotic resistance was established (Ben-Ami *et al.*, 2009) while in a recent study, age was not a substantial issue in defining the antibiotic resistance but gender plays a vital role in the sensitivity (Lee *et al.*, 2016). This study revealed that no antibiotic resistance significant variance in between the males and females ($P > .05$) which is also in agreement with a previous study by Ghadiri (Ghadiri *et al.*, 2014). In this present study, the maximum number (19 isolates) of urine carried 40% of MBLs producing gene which agreed with other study defined previously (Anago *et al.*, 2015). Furthermore, a study in Indonesia in which sputum was the highest carriage of these genes (Karuniawati *et al.*, 2013) and most (47.7%) resistant gene was isolated from pus swabs (Okoche *et al.*, 2015). More than 80% of the urinary tract infections (UTI) while female febrile UTIs 86.3% and male febrile UTIs 78.5% are due to *E. coli* (Lee *et al.*, 2016). Misappropriation of antibiotic is liable for the higher incidence of antibiotic resistance among bacteria (Neupane *et al.*, 2016). In this study, 65% *E. coli* isolates were multidrug resistant which are in agreement with a number of studies, 95.52% (Yadav *et al.*, 2015), 73.68% (Thakur *et al.*, 2013), 57.3% (Nairoukh *et al.*, 2018) and 40.7% isolates in Tehran (Ghadiri *et al.*, 2014). In this study, *E. coli* isolates exhibited variable rates of susceptibility and resistance to 10 different types of antibiotics. The highest percentage of resistant to CEX (89.1%) and cefepime (86.1%) agrees to 83.16% and 100% resistance respectively in Nepal (Jena *et al.*, 2017). The *E. coli* resistant rate of cefotaxime (84.4%) and ceftazidime (81.3%) (Chika *et al.*, 2016) are consistent with the results found in this study. In contrast, a higher resistant to ceftazidime (100%) and cefotaxime (100%) was observed (Jena *et al.*, 2017). However, among the carbapenem group of antibiotics, imipenem showed the least resistant (29.2%) correspond with data presented previously 17% (Jena *et al.*, 2017), 7.37% (Yadav *et al.*, 2015), 22% isolates of *E. coli* (Adwan *et al.*, 2014). In contrast, the higher resistant rate (82%) was observed in 2015 (Rezai *et al.*, 2015) while 100% isolates were imipenem resistant (Liang *et al.*, 2017) compared to this present study. Bacterial resistance due to metallo- -lactamase (MBL) production is increasing (Nazmul *et al.*, 2017). This study demonstrates screening for MBL producing imipenem resistant *E. coli* isolates (19) by three different phenotypic approaches differs in MHT (7, 36.8%), CDT (9, 47.4%) and DDST (10, 52.6%) for MBL positive strains. However, PCR results confirmed the occurrence of MBL genes in imipenem resistant *E. coli* 10/19 (52.6%) isolates same as by DDST method. Moreover, 70.0% isolates were MBL producers (Ghadiri *et al.*, 2014) which are agreeing to this study. This present study revealed a strong association between MBLs gene presence and MDR resistant *E. coli* (Figure 1). In addition, number of study observed a lower rate of 18.98% (Bora *et al.*, 2014), 12.5% (Jena *et al.*, 2017), 4.2% (Agbo and Eze, 2015) and 28.6% strains (Chika *et al.*, 2014) were MBL producing *E. coli*. The prevalence rate of MBLs 53% was observed in *E. coli* (Marie *et al.*, 2013) which is in agreement to this study and Mita and his colleagues (Wadekar *et al.*, 2013) has reported a lower prevalence rate of MBLs (13.4%). The variances between phenotypic and genotypic determination of MBL producing *Escherichia coli* isolates have been reported in earlier studies (Okoche et al., 2015). An increase acquisition of MBL genes such as blaIMP and blaVIM (Mohanam and Menon, 2017), (Y. Khosravi et al., 2010), (Salimi and Eftekhar, 2014), (Liew et al., 2018) was observed throughout the world. Among the MBL-encoding genes, blaVIM are the most prevalent worldwide, followed by blaIMP (Y. Khosravi et al., 2010; Salimi and Eftekhar, 2014) but this study revealed the prevalence of blaIMP is higher than blaVIM. The PCR amplification products of MBLs producing genes (blaIMP and blaVIM) was observed in this present study. Among the imipenem-resistant *Escherichia coli*, ten strains (52.6%) carried 1 or both MBL genes as shown in single PCR experiments (Figure- 5). Imipenem resistant gene blaIMP was observed in 7(70%) and blaVIM in 3 isolates (30%), of which 1 carried both genes. A recent study has revealed 8 imipenem resistant *E. coli* genotypically confirmed to harbor blaIMP gene by the multiplex PCR and while blaVIM was absent (Chika et al., 2016). In contrast, some of the recent studies showed both blaIMP and blaVIM genes were absent (Liang et al., 2017; Nairoukh et al., 2018).

## Conclusion

In conclusion, the prevalence rate of antibiotic resistance, both blaIMP and blaVIM type MBL producing *E. coli* strains is in alarming condition compared to other parts of the world. The hospital should highlight the clinically relevant strains for MBLs producing microorganisms and management of quality control infectious disease. The higher prevalence rate of antibiotic-resistant and -lactamase producing *E. coli* are associated risk factors demands epidemiological study, advance attention and concern to stop spreading of MBLs genes producing gram-negative bacteria in the community.

## Conflict of interest
None

## Acknowledgment
Authors like to appreciate each of the co-others for their unconditional support.